\documentclass[letterpaper, 10pt, conference]{ieeeconf}  

\IEEEoverridecommandlockouts                              

\usepackage{array}
\usepackage[caption=false,font=normalsize,labelfont=sf,textfont=sf]{subfig}
\usepackage{textcomp}
\usepackage{stfloats}
\usepackage{url}
\usepackage{verbatim}
\usepackage{graphicx}
\usepackage{epstopdf}
\usepackage{balance}

\usepackage{cite}
\usepackage{times}
\usepackage{multicol}
\usepackage[bookmarks=true]{hyperref}
\usepackage[dvipsnames, svgnames, x11names]{xcolor}
\usepackage{hyperref}
\usepackage{amsmath,amsfonts,amssymb}
\usepackage{siunitx}
\usepackage{standalone}
\usepackage{booktabs}
\usepackage[ruled,vlined,linesnumbered]{algorithm2e}
\usepackage{mdframed}
\usepackage{fancyvrb}
\usepackage{soul}
\usepackage{dsfont,mathabx}
\usepackage{array, booktabs}
\usepackage{threeparttable}
\usepackage{multirow}
\usepackage{algpseudocode}
\usepackage{amsthm}
\usepackage{float}
\usepackage{caption}
\usepackage{float}
\usepackage{subfloat}

\theoremstyle{definition}

\newtheorem{problem}{Problem}
\newtheorem{definition}{Definition}
\newtheorem{proposition}{Proposition}

\begin{document}
\title{\LARGE \bf Multi-agent Coordination Under Temporal Logic Tasks and\\
	Team-Wise Intermittent Communication
}

\author{Junjie Wang, Meng Guo and Zhongkui Li
	\thanks{The authors are with the State Key Laboratory
		for Turbulence and Complex Systems,
		Department of Mechanics and Engineering Science,
		College of Engineering, Peking University, Beijing, China.
		This work was supported by the National Key R\&D Program of China (2022ZD0116401, 2022ZD0116400), 
		the National Natural Science Foundation of China (U2241214, 62373008, T2121002).
		Corresponding author: \texttt{zhongkli@pku.edu.cn}.}
	}

\maketitle

\begin{abstract}
    Multi-agent systems outperform single agent in complex collaborative tasks.
    However, in large-scale scenarios, ensuring timely information exchange 
    during decentralized task execution remains a challenge. 
    This work presents an online decentralized coordination scheme for multi-agent systems 
    under complex local tasks and intermittent communication constraints. 
    Unlike existing strategies that enforce all-time or intermittent connectivity,
    our approach allows agents to join or leave communication networks 
    at aperiodic intervals, as deemed optimal by their online task execution. 
    This scheme concurrently determines local plans and refines the communication strategy, i.e., where and when to communicate as a team. 
    A decentralized potential game is modeled among agents,
    for which a Nash equilibrium is generated iteratively through online local search. 
    It guarantees local task completion and intermittent communication constraints.
    Extensive numerical simulations are conducted against several strong baselines.
    \end{abstract}


\section{Introduction}\label{sec:introduction}
Coordination for multi-agent systems (MAS) with temporal logics is gaining attention,
due to its intuitive framework to describe and coordinate complex tasks,
like surveillance and search-rescue missions
\cite{lin2014mission,sahin2019multirobot,buyukkocak2021planning}. 
Correct-by-design methods are proposed to synthesize both discrete plans and 
continuous control strategies for each agent.
Furthermore, inter-agent communication is crucial in facilitating 
real-time information exchange and collaborative coordination. 
However, existing approaches either assume fully-connected communication, 
or sacrifice task efficiency for communication. 
Instead, our work focuses on coordinating MAS under 
temporal logic tasks and communication constraints,
such that task performance can be guaranteed alongside sufficient team-wise communication.


Temporal logics, like linear temporal logic (LTL), 
serve as a formal language for complex high-level task description in MAS. 
Discrete plans are synthesized via off-the-shelf model-checking algorithms 
using workspace abstraction and temporal logic tasks
\cite{yan2021decentralized, schillinger2019hierarchical}. 
Two common formalisms exist for assigning temporal tasks.
One decomposes global tasks into local sub-tasks, 
assigned to individual agents in a top-down manner~\cite{chen2011formal, ulusoy2013optimality}. 
The other adopts a bottom-up approach, 
assigning individual tasks to agents~\cite{filippidis2012decentralized, guo2018multirobot}, 
and necessitating real-time coordination. 
Our work adopts the bottom-up formalism as each agent is initially assigned
specific local LTL tasks,
entailing collaboration and real-time information exchange.

As described earlier, communication is vital in MAS, 
enabling behavior coordination and real-time information sharing. 
Related literature categorizes communication into all-time and intermittent types. 
The former requires constant agent connectivity, 
often managing network connectivity through graph theory
\cite{oh2015survey,griparic2022consensus,khateri2019comparison,derbakova2011decentralized,
guo2016communication}. 
These approaches either preserve initial links~\cite{griparic2022consensus} 
or allow for link additions and removals~\cite{derbakova2011decentralized}
while ensuring network connectivity. 
However, they usually neglect local tasks for agents, 
focusing solely on maintaining connectivity. 
Our earlier work addresses both connectivity constraints and local temporal tasks
\cite{guo2016communication},
which however scarifies greatly the efficiency of task execution
as all agents move as a fixed topology to accomplish local tasks.
Therefore, the intermittent communication has emerged as an alternative solution
\cite{hollinger2012multirobot,kantaros2019temporal,aragues2020intermittent},
where the communication network is disconnected at most of the time.
Most recent methods propose decentralized pair-wise communication protocols
enabling communication among pairs or subgroups of the agents only when necessary. 
In other words the communication constraints are much more relaxed,
thus enhancing efficiency of task execution.
However, these methods usually impose fixed communication schedules 
\cite{kantaros2019temporal}, 
or communication at predetermined locations \cite{guo2018multirobot}. 
Furthermore, due to infrequent network connections, 
real-time information propagation faces delays,
impacting collaborative state estimation and data fusion processes.

This work introduces an online decentralized coordination scheme for MAS with 
complex local tasks and intermittent team-wise communication constraints. 
The agents are assigned complex local tasks expressed as LTL formulas involving 
their motion and actions.
Simultaneously, the team is required to maintain connectivity for at least $D_c$ 
during each time interval $T_c$, known as the $(T_c, D_c)$ constraint, 
to facilitate timely information exchange. 
To satisfy both local tasks and communication constraints, 
we formulate a constrained optimization problem 
encompassing task execution and communication strategies.
Our proposed method utilizes a decentralized potential game framework
for agents to optimize \emph{where} and \emph{when} to communicate as a team 
under the $(T_c, D_c)$ constraint. 
Through iterative online processes, a Nash-stable solution is generated recursively,
while achieving a balance between local task efficiency and intermittent communication.
The overall scheme is demonstrated by extensive numerical simulations
against several baselines.
The main contribution is two-fold:
i)~the formulation of a novel task coordination problem 
under team-wise intermittent communication constraints;
ii)~the proposed online decentralized coordination scheme 
for task execution and team-wise communication,
which ensures both the satisfaction of all local tasks
and a timely information propagation across the team.

\section{Preliminaries}\label{sec:preliminaries}
A~\textit{Linear Temporal Logic (LTL)} formula consists of a set of 
atomic propositions~$\Psi$ and several boolean and 
temporal operators~\cite{baier2008principles}: 
$\varphi::=\text{true}\,|\,\psi\,|\,\varphi_1\wedge\varphi_2\,|\,\neg\varphi\,|\,\bigcirc\varphi\,|\,\varphi_1\textsf{U}\varphi_2$,
where~$\neg$ (\textit{negation}), $\wedge$ (\textit{conjunction})
are standard boolean connectives and $\bigcirc$ (\textit{next}), $\textsf{U}$ (\textit{\textit{until}}) are temporal operators. 
Besides, the derivations of other useful operators, 
such as $\Diamond$~(\textit{eventually}), $\Box$~(\textit{always}), 
and the semantics of LTL are omitted here due to limited space. 
Particularly, \textit{syntactically co-safe LTL (sc-LTL)} is a subclass of LTL without the operator $\Box$, 
and requiring that the negation operator~$\neg$ appears only in front of atomic propositions~\cite{kupferman2001model}.
In addition, the words that satisfies an LTL formula~$\varphi$
over $\Psi$ can alternatively be captured through
a \textit{Nondeterministic B\"uchi Automaton (NBA)},
denoted by~$\mathcal{A}_\varphi=(Q,2^\Psi,\delta,Q_0,F)$,
where $Q$ is a finite set of states; $2^\Psi$ is a power set of all alphabets;
$\delta:Q\times 2^\Psi\times Q$ is a transition relation;
$Q_0$ and $F\subseteq Q$ is the set of initial and accepting states.

\section{Problem Formulation}\label{sec:problem}
Consider a team of~$N$ robotic agents coexisting in a 2D workspace with a cluster of obstacles.
Each agent~$i\in\mathcal{N}\triangleq\{1,\dots,N\}$ can only traverse within 
the free space~$\mathcal{Z}\in\mathbb{R}^2$ and employs the unicycle dynamics
\begin{equation}\label{eq:dynamics}
	\dot{x}_i=v_i \cos \left(\theta_i\right), \quad \dot{y}_i=v_i \sin \left(\theta_i\right), \quad \dot{\theta}_i=w_i,
\end{equation}
where~$p_i(t)=(x_i(t),\,y_i(t))\in\mathbb{R}^2$ and~$\theta_i(t)\in(-\pi,\pi]$ 
are the central position and orientation of agent~$i$ at time~$t\geq 0$.
\subsection{Local Task Specification}\label{subsec:agent-dynamic}
All agents can navigate within the free space~$\mathcal{Z}$ and perform various actions.
There is a set of regions of interest with different properties for each agent~$i\in\mathcal{N}$,
denoted by~$\Pi_i=\{\pi_{i,1},\cdots, \pi_{i,M_i}\}$,~$M_i>0$, 
which is assigned and known a priori.
Besides, each agent~$i$ is capable of several actions~$A_i=\{a_{i,0},a_{i,1},\dots,a_{i,L_i}\}$, $L_i>0$,
where~$a_{i,0}$ reflects none of actions.
The duration of each action is given by the function~$D_i:A_i\to\mathbb{R}_{\geq 0}$.
With a slight abuse of notations, we denote the set of atomic propositions 
by~$\Psi_i\triangleq\{\pi_{i,m}\wedge a_{i,l},\,\forall \pi_{i,m}\in\Pi_i,\, \, \forall a_{i,l}\in A_i\}$, 
over which a high-level local task for agent~$i$ can be specified as the LTL formula:
~$\varphi_i \triangleq \Box \Diamond (\varphi^{\texttt{s}}_i)$,
where~$\varphi^{\texttt{s}}_i$ is a co-safe LTL formula as defined in Sec.~\ref{sec:preliminaries}.
The task $\varphi^{\texttt{s}}_i$ can be satisfied in finite time,
while~$\varphi_i$ requires~$\varphi^{\texttt{s}}_i$ to be satisfied infinitely often.

\subsection{Communication Constraint}\label{subsec:communication}
To facilitate information exchange during task execution,
agents are required to communicate with others sufficiently often.
However, their communication is constrained by the relative distance,
i.e., two agents $i,j\in \mathcal{N}$ can only establish communication at time $t\geq0$ when~$d_{ij}(t) = \|p_i(t)-p_j(t)\|_2\leq R$,
where $R>0$ is the bounded communication range and uniform among all agents.
Consequently, the communication network is defined as a time-varying graph~$G(t) \triangleq (\mathcal{N},\, E(t))$,
where $E(t)\triangleq \{(i,\, j)\,|\,d_{ij}(t)\leq R\}$ is the time-varying edge set.
Thus, the team~$\mathcal{N}$ can perform communication at time $t\geq 0$,
if the underlying graph~$G(t)$ is connected.
Moreover, the team $\mathcal{N}$ is said to have 
\textit{sufficient communication} with $(T_c,\,D_c)$,
denoted by
\begin{equation}\label{eq:suf-com}
	\mathcal{N}\; \stackrel{\texttt{com}}{\sim}\;  (T_c,\,D_c),
\end{equation}
if there exists~$t'\in [rT_c,\, (r+1)T_c),\,\forall r\in\mathbb{N}$,
such that~$G(t)$ is connected, $\forall t\in [t',\, t'+D_c)$, 
where~$T_c>D_c>0$.


\subsection{Performance of Task Plan}\label{team-makespan}
Particularly, the complete plan~$\Gamma_i$ of agent~$i\in\mathcal{N}$ is an infinite
sequence of task states, combining the motion path with local actions 
performed at specified regions.
Moreover, let~$\mathbf{t}_i^\texttt{s}\triangleq t_{i,1}^\texttt{s}\cdots t_{i,k}^\texttt{s}\cdots$
be the time sequence when~$\varphi^\texttt{s}_i$ is satisfied 
during the execution of plan~$\Gamma_i$, i.e.,~$\varphi^{\texttt{s}}_i$ is performed 
for the $k$-th time during~$[t_{i,k}^\texttt{s}, \, t_{i,k+1}^\texttt{s})$
with~$t_{i,k+1}^\texttt{s} > t_{i,k}^\texttt{s} > 0$, $\forall k\in\mathbb{N}^+$.
Thus, the \emph{makespan} of agent~$i$ to perform co-safe LTL task is given by
$\xi_i\triangleq\textbf{max}_{k\in\mathbb{N}^+}\,
\{t_{i,k+1}^\texttt{s}-t_{i,k}^\texttt{s}\}$,
which is the maximum interval between two consecutive instances in~$\mathbf{t}_i^\texttt{s}$.
It actually measures the bottleneck efficiency of accomplishing the co-safe task,
by which, the performance of the whole team can be evaluated.


\subsection{Problem Statement}\label{subsec:problem}
\begin{problem}\label{pro:original}
	Consider a team of~$N$ agents with dynamics~\eqref{eq:dynamics} and 
	bounded communication range~$R$.
	Each agent~$i\in\mathcal{N}$ is assigned an LTL task $\varphi_i$.
	Design an integrated planning scheme for agents to synthesis their 
	complete plans~$\Gamma_i,\, i\in\mathcal{N}$, such that
	\begin{equation}\label{eq:op-pro}
		\begin{split}
			&\underset{\{\Gamma_i,\, i\in\mathcal{N}\}}{\textbf{min}}\ 
			\sum_{i\in\mathcal{N}}\,{\xi_i},\\
			s.t.\;
			&\varphi_i\, \text{is satisfied},\ \forall i\in\mathcal{N},\\
			&\mathcal{N}\; \stackrel{\texttt{com}}{\sim}\;  (T_c,\,D_c),
		\end{split}
	\end{equation}
	indicating that local tasks and communication constraints are satisfied,
	meanwhile the sum of makespan is minimized.
	\hfill $\blacksquare$
\end{problem}

\section{Integrated Planning of Local Task and Intermittent Communication}
\label{sec:solution}
In this work, we propose an online integrated plan and execution scheme,
which mainly contains three parts:
i)~the offline synthesis of local plan by each agent,
ii)~the online coordination of intermittent communication as a team,
and iii)~the hybrid execution of local plans and communication.

\subsection{Local Plan Synthesis}\label{sec:localplan}
To begin with, each agent $i\in\mathcal{N}$ synthesizes its local discrete plan 
that satisfies its LTL task~$\varphi_i$.
Since~$\varphi_i$ is independent among the agents, 
the local plans can be synthesized locally without coordination.
Similar to our previous work~\cite{guo2018multirobot}, the complete model of agent~$i$
that combines motion and actions among the regions of interest~$\Pi_i$
can be derived as a finite transition system, denoted by
$\mathcal{T}_{i}\triangleq (\Pi_i',\, \rightarrow_i',\, \Pi_{i,0}',\, \Psi_i,\, L_i,\, T_i')$,
where~$\Pi_i'=\Pi_i\times A_i$ is the set of composed states with
$\pi_{i,\tilde{m}}'=\langle\pi_{i,m}, a_{i,l}\rangle\in \Pi_i'$, 
$\rightarrow_{i} \subseteq \Pi_i' \times \Pi_i'$ is the transition relation
such that
$(\langle\pi_{i,m}, a_{i,l}\rangle,\,\langle\pi_{i,n}, a_{i,h}\rangle) 
\in \rightarrow_{i}$
if two conditions hold: i) there is a path from $\pi_{i,m}$ to $\pi_{i,n}$ and 
$a_{i,h}=a_{i,0}$,
or ii)~$\pi_{i,m}=\pi_{i,n}$ and~$a_{i,l},\, a_{i,h}\in A_i$;
$\Pi_{i,0}'\subset \Pi_i'$ contains the initial states;
$\Psi_i$ is the set of atomic propositions;
$L_i:\Pi_i'\to 2^{\Psi_i}$ is the labeling function,
and~$T_i':\to_i\to\mathbb{R}_{\geq 0}$ approximates the time cost of each transition,
measured by
\begin{equation}\label{eq:trans-time}
	T_i'(\pi_{i,\tilde{m}}',\,\pi_{i,\tilde{n}}')=T_i(\pi_{i,m},\pi_{i,n})+D_i(a_{i,h}),
\end{equation}
where~$\pi_{i,\tilde{m}}'=\langle\pi_{i,m}, a_{i,l}\rangle$
and~$\pi_{i,\tilde{n}}'=\langle\pi_{i,n}, a_{i,h}\rangle$,
$T_i(\pi_{i,m},\pi_{i,n})$ estimates the time cost of traversing the path from~$\pi_{i,m}$ to~$\pi_{i,n}$
by the turn-and-forward technique as in~\cite{guo2018multirobot},
and~$D_i(a_{i,h})$ is the duration time of action~$a_{i,h}$.

Given the complete model~$\mathcal{T}_i$,
the local plan of agent~$i$ can be synthesized via the automaton-based model-checking algorithms, see~\cite{baier2008principles}.
More precisely, the synchronized product automaton between~$\mathcal{T}_i$ and the NBA~$\mathcal{A}_{\varphi_i}$ associated with~$\varphi_i$ is constructed, denoted by~$\mathcal{P}_i=\mathcal{T}_i\times\mathcal{A}_{\varphi_i}$.
Then an accepting run of~$\mathcal{P}_i$
is synthesized with the prefix-suffix structure~$\tau_i=\tau_{i,\texttt{pre}} (\tau_{i,\texttt{suf}})^\omega$,
where~$\tau_{i,\texttt{pre}}=\pi_{i,0}'\pi_{i,1}'\cdots \pi_{i,k_i-1}'$ is the prefix that executed only once,
and~$\tau_{i,\texttt{suf}}=\pi_{i,k_i}' \cdots \pi_{i,K_i}',\,1\leq k_i\leq K_i$ is the suffix that repeated infinitely often,
which contains at least one state that is acceptable for local task~$\varphi_i$
Consequently, the makespan of agent~$i$ over infinite horizon is equivalent
to the total time cost of~$\tau_{i,\texttt{suf}}$, calculated 
by~$\xi_{i} =T_i'(\pi_{i,K_i}',\pi_{i,k_i}')+\sum_{k=k_i}^{K_i-1} T_i'(\pi_{i,k}',\pi_{i,k+1}')$.
Therefore, the optimal local plan~$\tau_i$,
can be found via a nest-Dijkstra shortest path algorithm,
i.e., a loop with minimum cost that contains at least one accepting state.
For more algorithmic and implementation details, please refer to~\cite{baier2008principles}.

However, the local plan are synthesized without considering the communication constraints in~\eqref{eq:suf-com}.
In other words, if these plans are executed blindly,
the agents can only communicate pair-wise \emph{by chance} 
rather than an intermittent full connection,
which would mostly violate the constraint of sufficient communication.

\subsection{Online Coordination of Intermittent Communication}\label{sec:coordination}
In this section, we design an approach for all agents
to schedule communication events including \emph{where} and \emph{when} to communicate as a team.
First, the Prob.~\ref{pro:original} is reformulated as a combinatorial optimization w.r.t. 
the communication events in finite horizon.
Under this formulation, a potential game is designed for which 
the concept of Nash equilibrium is introduced.
Then, a decentralized coordination scheme is proposed to 
find a communication strategy
that satisfies the Nash-stable condition.

\subsubsection{Problem Re-formulation}
For each agent~$i\in\mathcal{N}$,
the complete plan~$\Gamma_i$ is generated by scheduling
communication events properly into the local plan~$\tau_i$
synthesized in Sec~\ref{sec:localplan}.
The \emph{strategy} for agent to attend communication is 
a 2-tuple~$s_i\triangleq(z_{i,c},\,h_i)$,
where~$z_{i,c}\in\mathcal{Z}$ and~$h_i\in\mathbb{N}^+$ are 
the location and the index of subtask in plan~$\tau_i$,
which indicates that the agent will move to~$z_{i,c}$
for communication after the $h_i$-th task in local plan.
The \emph{team strategy} is the compilation of all local strategies,
denoted by~$S=\{s_i,\,i\in\mathcal{N}\}$.

To distinguish the communication events,
we use~$r\in\mathbb{N}$ as the round of communication,
where~$r=0$ indicates that all agents are connected at~$t=0$.
Therefore, the complete plan~$\Gamma_i$ can be split into 
the concatenating form, denoted by
$\Gamma_i\triangleq\tau_i^1\cdots\tau_i^r\cdots,\,r\in\mathbb{N}^+$,
where~$\tau_i^r\triangleq\tau_i[l_i^r\!:\!h_i^r]\, s_i^r$ 
combines a segment of local plan~$\tau_i$ 
from index~$l_i^r$ to $h_i^r$ with~$h_i^r\geq l_i^r\geq 1$,
and the communication strategy~$s_{i}^r$.
Particularly, there exists~$l_i^{r+1}=h_i^r+1,\,r\in\mathbb{N}^+$ and $l_i^1=1$.
Furthermore, the associated time sequence of plan~$\tau_i^r$ is generated 
by~$\mathbf{t}_i^{r}=t_{i,l_i^r} t_{i,l_i^r+1}\ldots t_{i,h_i^r}t_{i,s_i^r}$.
More importantly, the \emph{additional time} to attend communication using strategy~$s_i^r$ is computed by
\begin{equation}\label{eq:addition}
	\xi_{i,s_i^r}=\delta_{i,s_i^r} + (t_c^r-t_{i,s_i^r}),
\end{equation}
where~$\delta_{i,s_i^r}$ is the extra traveling time calculated 
by~$\delta_{i,s_i^r}=T_i(\pi_i^{h_i^r}, z_{i,c}^r) + T_i(z_{i,c}^r, \pi_i^{h_i^r\!+\!1})-T_i(\pi_i^{h_i^r}, \pi_i^{h_i^r+1})$,
where~$\pi_i^{h_i^r}$ and $\pi_i^{h_i^r\!+\!1}$ are the task locations before and after communication event;
while~$t_c^r=\textbf{max}_{i\in\mathcal{N}}\,\{t_{i,s_i^r}\}$ is the time when communication actually takes place,
and~$t_c^r-t_{i,s_i^r}$ is the waiting time for communication.
Here, we ignore the communication duration~$D_c$, for it is constant among agents.
Moreover, the communication topology is constructed by 
collecting all locations~$Z_c^r=\{z_{i,c}^r,\,i\in\mathcal{N}\}$ in the strategies.
It is denoted by~$G(Z_c^r) = (\mathcal{N}, \, E(Z_c^r))$, 
where~$E(Z_c^r)=\{(i,\,j)\,|\,\|z_{i,c}^r-z_{j,c}^r\|\leq R\}$ is the set of connected edges.
Based on the above descriptions, the considered Prob.~\ref{pro:original}
can be reformulated as follows.

\begin{problem}\label{pro:strategy}
	Design a coordination scheme for the team~$\mathcal{N}$ to find the optimal 
	strategy set~$S^r$, by solving the following combinatorial optimization problem 
	for each round~$r\in\mathbb{N}^+$: 
	\begin{equation}\label{eq:comb-opti}
		\begin{split}
			&\underset{S^r}{\textbf{min}}\, \sum_{i\in\mathcal{N}}\,\xi_{i,s_i^r},\\
			s.t.\;
			&G(Z_c^r)\,\text{is connected},\\
			&t_c^r\in[rT_c,\,(r+1)T_c),
		\end{split}
	\end{equation}
	which indicates that the sum of the additional time caused by communication is minimized, 
	while the constraints of intermittent communication are fulfilled.
	\hfill $\blacksquare$
\end{problem}

\subsubsection{Potential Game with Nash Equilibrium}
To balance the optimality and efficiency of the solution,
we model Prob.~\ref{pro:strategy} as a potential game among the agents as inspired 
by~\cite{monderer1996potential}.
Formally, a game is characterized by~$\Theta\triangleq (N,\,\mathcal{S}^N,\,u)$,
where~$N$ is the number of agents as players;
$\mathcal{S}^N=\mathcal{S}_1\times\cdots\times\mathcal{S}_N$ is the set of team strategies
defined over the strategy set~$\mathcal{S}_i$ of each agent~$i\in\mathcal{N}$;
and~$u:\mathcal{S}\times\mathcal{S}^{N-1}\to\mathbb{R}$ is the cost function.
Each agent aims to minimize the cost~$u_i$ by changing its strategy~$s_i$ within
the strategy set~$\mathcal{S}_i$.
Then, an exact potential game is defined as follows.
\begin{definition}\label{def:game}
	A game~$\Theta=(N,\,\mathcal{S},\,u)$ is an exact \emph{potential game}, 
	if there is a potential function~$\Phi:\mathcal{S}^N\to\mathbb{R}$,
	such that
	\begin{equation}
		u(s_i',\,S_{-i}) - u(s_i,\,S_{-i}) = \Phi(S') - \Phi(S),
	\end{equation}
	$\forall s_i,\,s_i'\in\mathcal{S}_i$, $\forall S_{-i}\subset\mathcal{S}^{N-1}$, 
	$\forall i\in\mathcal{N}$, where~$S_{-i}=S/\{s_i\}$ and~$S'=S_{-i}\cup\{s_i'\}$.
	\hfill $\blacksquare$
\end{definition}
Here, we set the potential function as the sum of the additional time defined 
by~\eqref{eq:addition},
i.e.,~$\Phi(S)\triangleq \sum_{i\in\mathcal{N}}\,\xi_{i,s_i}$,
and the cost function for each agent~$i\in\mathcal{N}$ 
as~$u(s_i,\,S_{-i})\triangleq \delta_{i,s_i}+Nt_c-t_{i,s_i}$.
Then, Prob.~\ref{pro:strategy} can be related to an exact potential game.
Particularly, the change in cost if agent~$i$ switches from strategy~$s_i$ 
to~$s_i'$ is measured by
\begin{equation}\label{eq:change-value}
	\begin{split}
		\sigma_{i,S_{-i}}^{s_i \to s_i'} \triangleq  u_i(s_i',\,S_{-i}) - u_i(s_i,\,S_{-i}),\\
	\end{split}
\end{equation}
which is equivalent to the change in team potential.
More importantly, the concept of Nash equilibrium for game~$\Theta$ is defined as below.
\begin{definition}\label{def:nash}
	A team strategy~$S\in\mathcal{S}^N$ is a \textit{Nash equilibrium}, if and only if,
	there does not exist any agent~$i\in\mathcal{N}$ that can reduce its cost
	by unilaterally switching to a new strategy~$s_i'\in\mathcal{S}_i$ from~$s_i$,
	while others retain the same strategies, 
	i.e.,~$\sigma_{i,S}^{s_i \to s_i'}\geq 0$,~$\forall s_i'\in\mathcal{S}_i$,~$\forall i\in\mathcal{N}$.
	\hfill $\blacksquare$
\end{definition}

Note that the potential game has two vital properties: 
i)~at least one Nash equilibrium exists,
and ii)~finite improvements can be made to the potential.
Ultimately, our goal is to find a Nash equilibrium strategy for Prob.~\ref{pro:strategy},
under which no agent can update its strategy individually to 
reduce the sum of additional time caused by communication.


\subsubsection{Decentralized Coordination Scheme}
In this section, an online decentralized coordination scheme
for the potential game is proposed,
which mainly contains two steps: first find the initial team strategy 
with the minimum potential among a class of special strategies, 
where all agents gather together at the same location;
then the initial team strategy is refined iteratively 
through a local search algorithm until it converges to a Nash equilibrium.

\textbf{Optimization of Initial Strategy}: 
Initially, each agent~$i$ prepares its tentative local plan in the finite horizon
according to the results of last round.
Particularly, the associated team strategy~$S_\texttt{N}$ is featured 
with the same location~$z_c\in\mathcal{Z}$ to communicate.
Then the initial strategy for round~$r\in\mathbb{N}^+$ is optimized by
\begin{equation}\label{eq:opt-node}
	S_\texttt{N}^\star=\underset{S_\texttt{N}\in\mathcal{S}}{\textbf{argmin}}\;
	\left\{\Phi(S_\texttt{N})\,|\,t_c\in[rT_c,\,(r+1)T_c)\right\},
\end{equation}
where the graph connectivity constraint in~\eqref{eq:comb-opti} is relaxed.
Eventually, the optimal team strategy~$S_\texttt{N}^\star$ can 
be determined by iterating all discrete nodes within workspace~$\mathcal{Z}$,
and the corresponding time set~$T_\texttt{N}^\star=\{t_{i,s_i},i\in\mathcal{N}\}$ 
is generated by collecting the arrival time of each agent~$i$ by strategy~$s_i$.

\textbf{Nash Equilibrium Strategy}:
As summarized in Alg.~\ref{alg:GNE}, an anytime algorithm is proposed to improve
the initial strategy for the game~$\Theta$ iteratively by local search,
until it converges to a Nash equilibrium or it exceeds 
a maximum iteration~$K\in\mathbb{N}^+$.
More specifically, within each iteration~$k<K$, 
agent~$i\in\mathcal{N}$ initially assembles a set of regions 
encompassing finite locations that 
maintain the connectivity of the communication topology,
denoted by
\begin{equation}
	\Upsilon_i(Z_c^k)=\{z\in\mathcal{B}(z_{i,c}^k,R)\,|\,G(Z_{c,-i}^k,z)\,\text{is connected}\},
\end{equation}
where~$\mathcal{B}(z_{i,c}^k,\,R)$ is a circular region around~$z_{i,c}^k$ within range~$R$, 
and $Z_{c,-i}^k=Z_{c}^k/\{z_{i,c}^k\}$.
Moreover, for any location~$z_i'\in\Upsilon_i(Z_c^k)$, 
the maximal decrease of the potential induced by agent~$i$ is given by
${\textbf{min}}\,\{\sigma_i\,|\,rT_c \leq t_c< (r+1)T_c\}$,
where~$\sigma_i=\sigma_{i,S^k_{-i}}^{s_i\to s_i'}$ is the change caused
by agent~$i$ switching from strategy~$s_i$ to~$s_i'=(z_i',\,h_i')$ 
under the current team strategy~$S^k$.

Therefore, the best strategy for agent~$i$ is generated by reducing the change $\sigma_i$ 
in a greedy way,
i.e., traversing the region set~$\Upsilon_i(Z_c^k)$ to find the strategy 
with minimal change $\sigma_i$ (Line~\ref{alg-line:search}-\ref{alg-line:reset}).
In addition, each agent compares its minimal~$\sigma_{i}^\star$ with its neighbors 
to obtain the minimal value~$\sigma_{i^\star}^\star$.
If it is no less than zero, then the current team strategy is a Nash equilibrium by 
Definition~\ref{def:nash} (Line~\ref{alg-line:zero}-\ref{alg-line:nash});
if not, the team strategy is updated by the optimal team strategy~$S_{i^\star}$, 
together with the time set~$T_{i^\star}$ (Line~\ref{alg-line:optimal-value}).
Through enough iterations, it converges to a Nash equilibrium, 
where the potential cannot be reduced any more.
The above process is repeated iteratively until it converges to a Nash equilibrium,
or exceeds the maximum iteration~$K$.
Note that the proposed algorithm can return a valid team strategy
when interrupted at any time, 
and the potential, i.e., the summation of additional time is decreased monotonically.
\begin{algorithm}[t]
	\caption{Decentralized Coordination Algorithm.}
	\label{alg:GNE}
	\LinesNumbered
	\KwIn {Team strategy~$S_\texttt{N}$ and time set~$T_\texttt{N}$;}
	\KwOut {Nash equilibrium team strategy~$S_\texttt{NE}$;}
	Initialize~$k\leftarrow 0$ and $S^0,\,T^0\leftarrow S_\texttt{N},\,T_\texttt{N}$\;
	\While{$k < K$}{
		\ForEach{$i\in\mathcal{N}$}{\label{alg-line:agent}
			Construct region set~$\Upsilon_i(Z_c^k)$\;\label{alg-line:region}
			Set~$\sigma_i^\star\leftarrow0$\;\label{alg-line:search}
			\For{$z_i'\in\Upsilon_i(Z_c^k)$}{
				$\sigma_i',\,h_i'\leftarrow\underset{h_i}{\textbf{min}}\,\{\sigma_i\,|\,rT_c \leq t_c< (r+1)T_c\}$\;
				\If{$\sigma_i'< \sigma_i^\star$}{
					Reset~$\sigma_i^\star\leftarrow \sigma_i'$\;
					$S_i,\,T_i \leftarrow S_{-i}^k\cup\{(z_i',\,h_i')\},\,T_{-i}^k\cup\{t_{i,z_i'}^{h_i'}\}$\;\label{alg-line:reset}
				}
			}
		}
		Determine~$i^\star=\underset{i\in\mathcal{N}}{\textbf{argmin}}\;\{\sigma_i^\star\}$\;\label{alg-line:compare} 
		\If{$\sigma_{i^\star}^\star\geq 0$}{\label{alg-line:zero}
			$\textbf{return}$ {$S_\texttt{NE}\leftarrow S^{k}$}\;\label{alg-line:nash}
		}
		$k\leftarrow k+1$, then $S^{k},\,T^{k} \leftarrow S_{i^\star},\,T_{i^\star}$\label{alg-line:optimal-value}\;
	}
	$\textbf{Return}$ {$S^K$}.
\end{algorithm}

\subsection{Hybrid Execution}
The complete plan~$\Gamma_i,\,i\in\mathcal{N}$ is synthetically  executed 
by interleaving between local tasks and communication events for each agent.
Initially, agents are located within a connected topology, 
each agent~$i$ first synthesizes the local plan~$\tau_i$ 
as described in Sec.~\ref{sec:localplan},
then coordinates with other agents to determine the communication location 
and time for the next round~$r=1$ using the scheme designed in Sec.~\ref{sec:coordination}.
Afterwards agents move and execute actions independently 
executing their local plans without communication, 
until they reach the agreed locations at the expected time.
However, due to the different efficiencies of performing local tasks, 
the agents mostly reach the agreed locations at different times.
So the agent that arrives early is responsible to wait for the later agents,
and the communication will happen eventually when the last agent arrives.
Since communication events always happen at the time when all agents gather together, 
the delays caused by different arrival times of agents do not propagate to the future round.

Once all agents have successfully reached the designated locations~$z_{i,c}^r,\,i\in\mathcal{N}$, 
a communication topology naturally takes shape.
Within this topology, the accumulated information can be comprehensively shared among all agents.
Subsequently, each agent prepares its preliminary local plan for 
the ensuing round $r$ building upon the tasks previously executed, 
and collaborates to formulate the subsequent communication plan $\Gamma_i^r$.
The above process continues iteratively, encompassing rounds from $r$ to $r+1$,
ensuring the fulfillment of local tasks and 
facilitating intermittent team-wise communication.

\begin{proposition}
	Starting from a connected communication topology,
	the hybrid execution framework ensures that each agent~$i\in\mathcal{N}$ can satisfy its local task~$\varphi_i$ and the team communication can happen sufficiently by~$(T_c,\,D_c)$.
	\hfill $\blacksquare$
\end{proposition}
\begin{proof}
	First, the correctness of the local plan~$\tau_i$ by each agent~$i\in\mathcal{N}$ is guaranteed by the model-checking algorithm~\cite{baier2008principles}.
	Consider the sequence of task states in complete plan~$\Gamma_i$ is equivalent to the local plan~$\tau_i$,
	when ignoring communication states, 
	so the local task~$\varphi_i$ can be satisfied by executing the complete plan~$\Gamma_i$.
	Moreover, the communication events are determined strictly within 
	time constraints~$(T_c,\,D_c)$ according to Prob.~\ref{pro:strategy},
	and due to the waiting mechanism, each agent will wait for communication 
	to happen before performing their individual tasks.
	Therefore the communication is guaranteed to iteratively happen at the time~$t_c^r=\textbf{max}_{i\in\mathcal{N}}\,\{t_{i,c}^r\}\in[rT_c,\,(r+1)T_c),\, r\in\mathbb{N}^+$, 
	by following the hybrid execution framework.
\end{proof}

\section{Numerical Simulation}\label{sec:experiments}
This section contains the numerical simulation
compared with several strong baselines.
The proposed approach is implemented in Python3
and run on a workstation with 12-core Intel Conroe CPU.
The complete simulation video can be found in
\href{http://y2u.be/t2PUp2_NdrU}{http://y2u.be/t2PUp2\_NdrU}.

\subsection{Workspace and Task Description}\label{}
\begin{figure}[t]
	\centering
	\subfloat{
		\label{subfig:workspace}
		\includegraphics[width=0.44\linewidth]{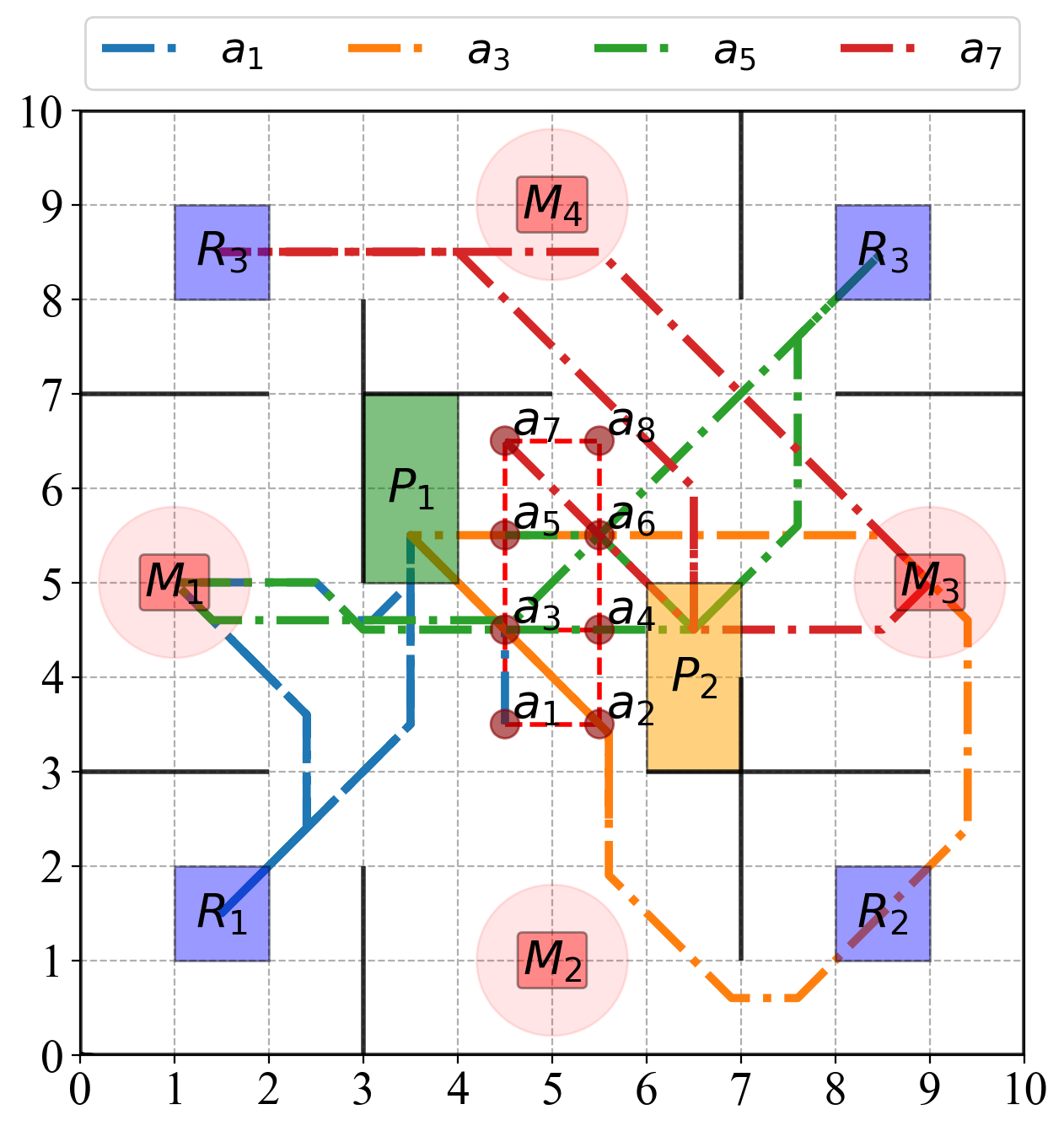}
	}
	\subfloat{
		\label{subfig:heatmap}
		\includegraphics[width=0.48\linewidth]{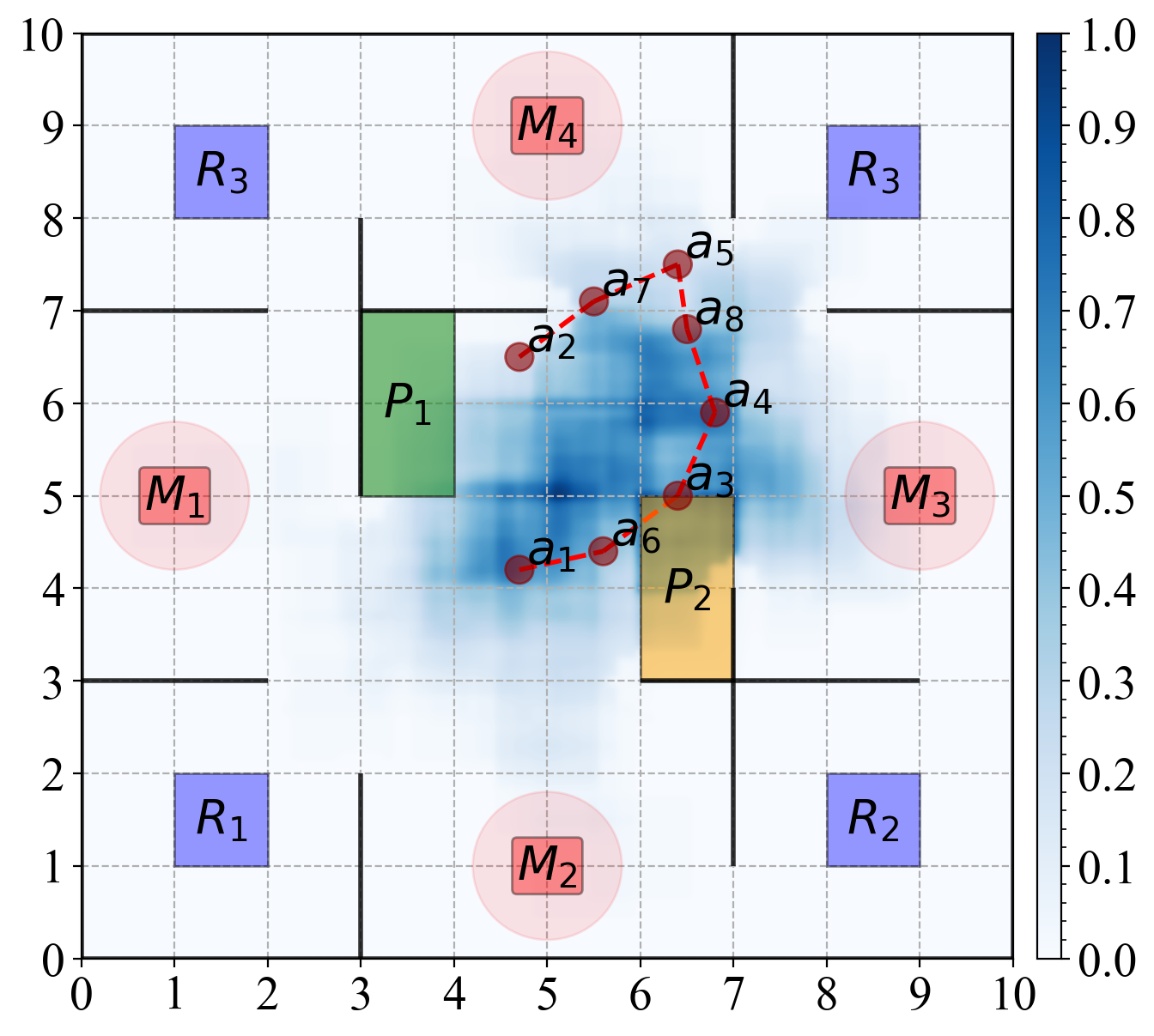}
	}
	\\
	\subfloat{
		\label{subfig:topologies}
		\includegraphics[width=0.94\linewidth]{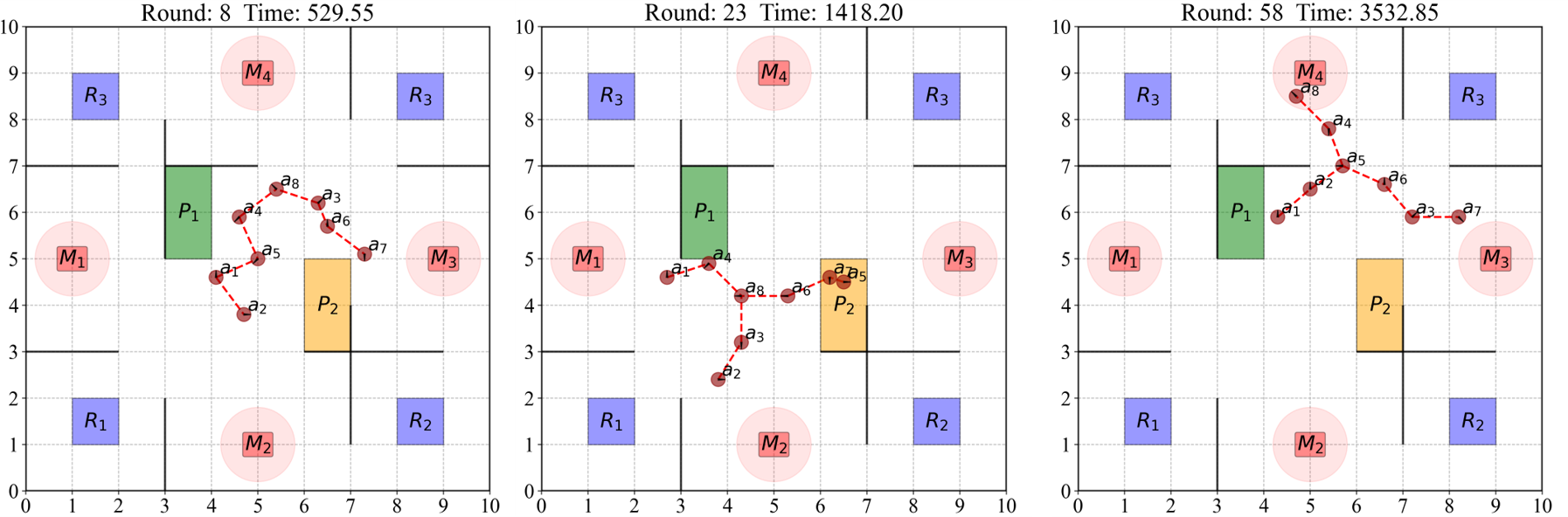}
	}
	\captionsetup{font={small}}
	\caption{(\textbf{Top-left}) Workspace abstraction and 
		the initial locations of agents,
		together with several local paths of agents~$1,3,5,7$;
		(\textbf{Top-right}) The topology of round~$r=1$, 
		and the distribution of topologies shown as a heat-map;
		(\textbf{Bottom}) The instances of topology in round~$r=8,\,23,\,58$.
	}
	\label{fig:local-plan}
\end{figure}
Consider a team of $N=8$~agents that coexist within a $10\,\text{m}\times10\,\text{m}$ workspace,
which is partitioned into cells as shown in Fig.~\ref{subfig:workspace}.
All agents follow the unicycle dynamics by~\eqref{eq:dynamics} with the reference linear
velocity~$v=1.0\,\text{m/s}$ and angular velocity~$\omega=1.5\,\text{rad/s}$.
Their communication ranges are uniformly set to $1.0\,\text{m}$.
Each agent is responsible for transporting resources from the production line~$P_{1,2}$
to the corresponding repository~$R_{1,\ldots,4}$,
monitoring air pollution index~(API) of the plant at different location~$M_{1,\ldots,4}$.
More specifically, the local task specifications for each agent are defined as~$\varphi_i = \Box\Diamond((\pi_i^\texttt{p}\wedge a_i^\texttt{c})\wedge\Diamond((\pi_i^\texttt{r}\wedge a_i^\texttt{u})\wedge\Diamond(\pi_i^\texttt{m}\wedge a_i^\texttt{m})))$,
which requires agent~$i$ to collect resources ($a_i^\texttt{c}$) at the production line~$\pi_i^\texttt{p}$,
then unload ($a_i^\texttt{u}$) the collected resources to the corresponding repository~$\pi_i^\texttt{r}$,
and finally move to the measurement station~$\pi_i^\texttt{m}$ to monitor ($a_i^\texttt{m}$) the current API, infinitely often.
See the simulation video for the exact task assignments.
The time cost for executing each action is set to $4$s uniformly.
Moreover, agents are encouraged to communicate sufficiently as a team with~$(T_c,D_c)$
as defined in~\eqref{eq:suf-com} to share their measured value in different stations.

\subsection{Simulation Results}\label{sec:exp-results}
This section mainly presents the results for the integrated planning scheme designed in Sec.~\ref{sec:solution}.
To begin with, agents start from initially-connected positions as shown in Fig.~\ref{subfig:workspace},
and the communication constraint $(T_c,\,D_c)$ in~\eqref{eq:suf-com} is set
to~$T_c=60.0$s and~$D_c=5.0$s.
Prepared with local plans, agents first coordinate with each other to find an initial strategy
with the average additional time $10.48$s for the 1st round,
then refine it by local search and converge to a Nash equilibrium strategy
eventually with $8.72$s using Alg.~\ref{alg:GNE}.
As shown in Fig.~\ref{subfig:heatmap}, the resulting communication topology is drastically different
from the initial one, both in location and topology.
The same procedure is simulated for $100$ rounds,
it takes~$1.323$s on average to generate the Nash equilibrium strategy, and the average additional time caused by communication is $7.72$s.
Besides, the topology distribution is shown as a heat-map in
Fig.~\ref{subfig:heatmap} and several intermediate topologies are in Fig.~\ref{subfig:topologies}.
Moreover, the final schedule of local tasks and communication events
indicates that the local tasks of each agent are satisfied,
and communication events occurs~$5.0$s every~$60.0$s.
\subsection{Comparisons with Baselines}\label{subsec:comparision}
\begin{figure}[t]
	\centering
	\includegraphics[scale=0.30]{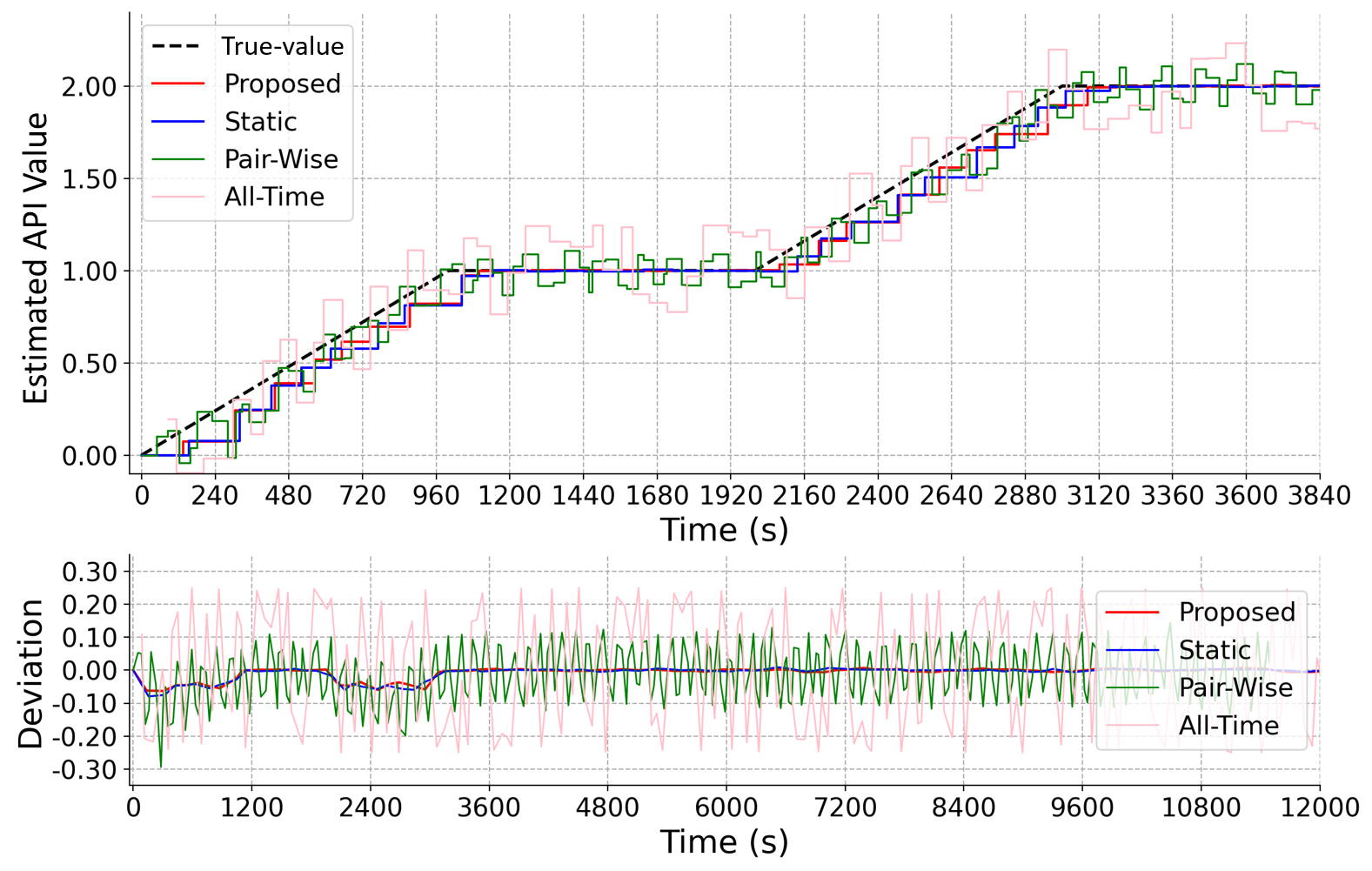}
	\captionsetup{font={small}}
	\caption{(\textbf{Top}) Evolution of the estimated API w.r.t. 
		the true value (in black dotted line)
		under different methods;
		(\textbf{Bottom}) Deviation of estimated API, which is the difference 
			between the estimated value and the true value at the same time instant.
	}
	\label{fig:API}
\end{figure}

The proposed scheme is compared against the following
three different baselines that are commonly used in multi-agent systems.
\textbf{Static}: the communication topology remains
static thus the same as the initial topology.
\textbf{Pair-Wise}: the agents communicate in pairs whenever possible and
the next communication round is optimized only between two agents,
see~\cite{kantaros2019temporal} for more detail.
\textbf{All-Time}: the agents are connected at all time and take turns to
execute its local task in a round-robin fashion,
as used in our earlier work~\cite{guo2016communication}.
The actual API in the plant follows a piecewise-linear function~$\nu(t)$
painted in black dotted line as shown in Fig.~\ref{fig:API}.
However, due to airflows in the plant,
the API values measured by the agents at stations~$M_{1,\ldots,4}$ drift with time.
In particular, it is assumed that an additional sinusoid noise is added to~$\nu(t)$
with magnitude 1 and frequency~$\pi/6000$.
Whenever the agents communicate, the consensus protocol~\cite{olfati2007consensus}
is followed to compute the average API.
All methods are simulated for the same duration, during which the
progress of task execution, communication events and locally estimated value of API
are stored.
The final results are summarized in Fig.~\ref{fig:API}.
It can be seen that the proposed method has the most accurate estimate of~$\nu(t)$
as each agent has the smallest deviation to the true value,
i.e., the maximum deviation is $0.82$ for our approach, $0.85$ for the static method.
While the estimated values of pair-wise and all-time communication 
have very large fluctuations compared to the true value,
and the latter has the largest deviation.
Moreover, the average additional time for agents to attend communication
by our proposed scheme is $13.15$s, while the time of static approach
is $20.04$s, which means that our proposed scheme is more time-saving.

\section{Conclusion}\label{sec:conclusion}
This work has proposed an online decentralized coordination scheme for multi-agent systems,
which features the co-design of the local plans and communication strategy,
subject to individual temporal tasks and communication constraints~$(T_c,D_c)$.
Different from all-time connected and pair-wise intermittent communication,
our scheme allows the agents to join and leave 
different communication topologies at aperiodic intervals, 
which ensures an efficient and timely data spreading across the team,
and also the satisfaction of local tasks.
Future research work includes online adaptive planning 
to dynamic environments and failure tolerance.


\bibliographystyle{IEEEtran}
\bibliography{contents/references}

\end{document}